\documentstyle[pra,aps,epsfig]{revtex}
\begin{document}
\draft

\draft\twocolumn[\hsize\textwidth\columnwidth\hsize\csname
@twocolumnfalse\endcsname

\title{Trends in Resonance Energy Shifts and Decay Rates for Bose Condensates
\\in a Harmonic Trap}
\author{Kunal Das and Thomas Bergeman}
\address{Department of Physics and Astronomy, SUNY, Stony Brook, NY 11794-3800
\\and Institute for Theoretical Atomic and Molecular Physics,
Harvard-Smithsonian Center for Astrophysics, 60 Garden Street,
Cambridge, MA 02138}
\date{\today }
\maketitle
\begin{abstract}
This is a study of quasi-discrete Bogoliubov quasi-particles in a spherically
symmetric harmonic trap.  We first evaluate analytically the aymptotic energy
shifts of the high energy modes and find them to have $1/\sqrt{n}$ dependence
on the number of radial nodes, $n$, consistent with earlier semiclassical
discussions.  To address the question of the widths or decay rates, $\gamma$,
we attempt to clarify previous discussions by deriving an implicit equation
for $\gamma$ from an assumption of exponential decay.  Numerically, we study
the trends in the behavior of the widths as a function of temperature, energy,
particle number and scattering lengths. In particular, we find that the width
due to Landau decay rises rapidly at low $n$ and then declines, while
the Beliaev decay rate rises slowly with $n$.  As temperature $\rightarrow 0$,
Beliaev decay reaches a constant ($>0$ for $n>0$), while the Landau decay rate
goes to zero.  The decay rate, $\gamma$, is approximately linear in the
$s$-wave scattering length.
\end{abstract}
]

\section{introduction}

In many finite quantum systems, such as atoms, nuclei, and molecules, an
extensive structure of excited states plays an important role in dynamical
situations and for diagnostics.  A homogeneous quantum fluid, such as
liquid $^4$He, exhibits a few excitation modes whose energy as a function of
wavenumber and temperature has been intensively studied \cite{GriffinB}.
Harmonically trapped, dilute alkali atom Bose condensates have recently
been found to have at least two prominent excitation modes
\cite{Jin1,Jin2,Mewes,StKurn} and theoretically there is the full manifold
of Bogoliubov quasi-particle modes.  In calculating thermal averages, one
can perform a sum over such discrete quasi-particle modes and obtain results
comparable to the more commonly used semiclassical integral over momentum
space \cite{GPS}. This raises the issue of the width of higher-lying
resonances, which is accentuated by recent observations of surface
modes \cite{Onofrio} and a chance observation of a longitudinal
mode \cite{Ketterle4}.  In other words, if experimental conditions are such
that higher modes {\it can} be excited, under what circumstances are their
widths narrow enough that they can be resolved?  Could collective or
single-particle excited states of Bose condensates become research tools
as useful as Rydberg states of atoms?

We first focus on the energies of excited modes. Analytic results
have been obtained \cite{Stringari1,WuGriffin,Ohberg} most notably
from the hydrodynamic approximation, which involves various
approximations to the kinetic energy. For highly excited states,
these approximations break down. Without making the hydrodynamic
approximation, Refs. \cite{HZG,You1,HZ,HDB} used Bogoliubov theory to
numerically compute low-lying excitations. High-lying levels were
calculated numerically by You and Walsworth \cite{Walsworth}.
Semiclassical methods have enabled Csord\'{a}s {\it et al.}
\cite{CGS1,CGSWKB} to obtain precise expressions for the shifts
(relative to bare harmonic oscillator states) of high lying
excitations in a spherically symmetric trap. The more general case
of cylindrical symmetry was found to lead to chaotic motion.  In
this paper, we present a very simple derivation of the lowest
order approximation to the energy shifts, somewhat in the spirit
of the quantum defect parameterization of excited Rydberg levels
of atoms, and we also present numerical results from
Bogoliubov-Popov theory.

Since the first experimental observations of resonance excitation and decay
in harmonic traps \cite{Jin1,Jin2,Mewes,StKurn}, a number of theoretical
approaches have been presented on the question of resonance width or decay
\cite{Liu,PS,Fedichev1,Fedichev2,Kavoulakis,Bene,G1,G2,Stoof1,Stoof2,Reidl2,GP,Williams}.
Much of this effort has been directed toward the JILA data, although more
recently Landau and also Beliaev damping of a ``scissors'' mode in a TOP
trap has been observed at Oxford \cite{Hodby}.
Arguably the most successful approach in explaining the JILA data has been
kinetic theory with Gaussian Ansatz for the condensate and excited mode
plus collision integrals for the condensate-noncondensate interaction
\cite{Stoof1,Stoof2}.
It was concluded that at a certain temperature the observed mode changed
character from out-of-phase to in-phase oscillation of the condensate and
the thermal atoms, and with this interpretation the calculations matched the
experimental data
quite well.  Good agreement has been offered also by calculations with
a stochastic sampling of classical orbits corresponding to quasi-particle
quantum states \cite{Fedichev1} and more recently with the dielectric formalism
\cite{Reidl2}.  In the present work, we will not consider again this
experimental data but instead address the question of the widths of
high-lying excitations in a spherically
symmetric trap.  Our approach makes use of accurate (numerical)
quasi-particle modes, and is valid when the perturbation of the system
is small. This approach has the advantage of making it possible to
consider explicitly the decay channels for Landau and Beliaev
decay processes.  We will present numerical results for decay widths
as a function of atom number, interaction strength and temperature.

Expressions for Landau and Beliaev decay of resonance excitations of trapped
Bose gases have been presented by Pitaevskii and Stringari \cite{PS}, by
Guillemas and Pitaevskii \cite{GP}, and by Giorgini \cite{G1,G2}.
Effectively, we adopt the expressions given in these papers, but provide
additional justification. Our discussion illuminates some of the many-body
aspects and the analogy with Weisskopf-Wigner theory of radiative decay.
One difficulty with previous expressions is that the terms in the sum over
decay modes involved a delta function in energy, which, if interpreted
exactly, will tend to be null for discrete quasi-particle states.
By introducing the Weisskopf-Wigner {\it Ansatz}
of an exponentially decaying initial state, we obtain decay
terms with a Lorentz profile factor in place of the energy delta function.

In this paper, our numerical results are intended not primarily to
establish precise values for the energy shifts and the damping
widths, but instead to seek patterns and trends as
a function of various parameters. We consider only a
spherically symmetric trapping potential. An outline of the paper
is as follows. In Sec. II, we review the basic equations of
Bogoliubov-Popov theory. In Sec. III, we use time independent
perturbation theory to establish analytically, within the
Thomas-Fermi approximation, the asymptotic behaviour of the energy
shifts.  In Sec. IV.A we review the
basic theory of exponential decay in time-dependent perturbation
theory and then proceed to derive an implicit equation for the
decay width. In Sec. IV.B, the lowest order interaction terms that lead
to damping of an oscillating condensate are derived.  Sec. IV.C uses
these results in conjunction with
the methods developed in Sec. IV.A to derive an expression for the
decay rate of an oscillating condensate. Section V
presents numerical methods and results of numerical
computation for the width and its dependence on various physical
parameters.

\section{Hamiltonian and Bogoliubov Equations}

The second-quantized grand-canonical Hamiltonian for a weakly
interacting Bose gas in a spherical harmonic trap is \begin{eqnarray}
\label{Kdef} \hat{K}=\int d{\bf r}\,
\hat{\psi}^{\dag}\left(H_{\rm{ho}}-\mu\right)\hat{\psi}+
\frac{g}{2}\int d{\bf r}\
\hat{\psi}^{\dagger}\hat{\psi}^{\dagger}\hat{\psi}\hat{\psi},
\end{eqnarray}
where $\mu$ is the chemical potential and \begin{eqnarray}
H_{\rm{ho}}=\frac{-\hbar^{2}\nabla^{2}}{2m}+\frac{1}{2}m\omega_{\rm{ho}}^{2}
r^{2}\hspace{2mm}; \hspace{1cm} g=4\pi\hbar^{2}a/m.
\end{eqnarray}
with $a$ the s-wave scattering length and $\omega_{\rm{ho}}/2 \pi$
the harmonic frequency of the trap. The field operators are
decomposed in the usual way into a macroscopic condensate
wavefunction and a noncondensate operator
\begin{equation}\label{Bogdec}
\hat{\psi}({\bf r},t)\simeq \langle\hat{\psi}({\bf
r},t)\rangle+\hat{\phi}({\bf r},t) =\Phi({\bf
r},t)+\hat{\phi}({\bf r},t).
\end{equation}
The temporal dependence of the condensate wavefunction is
attributed entirely to a small amplitude oscillation (fluctuation)
of the condensate about its equilibrium value $\Phi_{0}({\bf r})$
\begin{eqnarray}\label{flucsep} \Phi({\bf r},t)= \Phi_{0}({\bf
r})+\delta\hat{\Phi}({\bf r},t).
\end{eqnarray}

These substitutions are made and the resulting expressions
linearized with respect to the fluctuations. Subsequently
equilibrium mean fields $n_{T}({\bf r}) = \langle
\hat{\phi}^{\dagger}\hat{\phi}\rangle$ and $m_{T}({\bf r}) =
\langle \hat{\phi} \hat{\phi} \rangle$ are introduced to reduce
the Hamiltonian to a quadratic form in the noncondensate
operators. If $\Phi_{0}({\bf r})$ satisfies the Gross-Pitaevskii
equation \begin{eqnarray} \left[H_{\rm{ho}}-\mu
+g(|\Phi_{0}|^2+2n_{T})\right]\Phi_{0}=0, \label{GPE} \end{eqnarray} the
first order terms in $\hat{\phi}$ cancel out. The present work
does not deal with the regime near the critical temperature,
$T_{c}$, and thus the Popov approximation will be adequate and so
terms in ${m}_{T}$ are neglected.

If the coupling between the fluctuations and the thermal operators
are left out, the rest of the Hamiltonian is diagonalized by the
Bogoliubov transformation:
\begin{eqnarray} \label{Bog} \hat{\phi}({\bf
r},t)=\sum_{k}\ [\ u_{k}({\bf r})\hat{\alpha}_{k}(t)\ +\
v^{*}_{k}({\bf r}) \hat{\alpha}^{\dagger}_{k}(t)]\nonumber\\
\hat{\phi}^{\dagger}({\bf r},t)=\sum_{k} [u^{*}_{k}({\bf
r})\hat{\alpha}^{\dagger}_{k}(t)\ +\ v_{k}({\bf r})
\hat{\alpha}_{k}(t)],
\end{eqnarray}
provided that the functions $u_{k}({\bf r})$ and $v_{k}({\bf r})$
obey the Bogoliubov equations, \begin{eqnarray} \label{Bog1}
\hspace{2mm} \left[H_{\rm{ho}}-\mu+2
g(|\Phi_{0}|^2+n_{T})\right]u_{k}({\bf r})+g(|\Phi_{0}|^{2})
v_{k}({\bf r})\nonumber\\ = \epsilon_{k} u_{k}({\bf r}) \nonumber
\\ -\left[H_{\rm{ho}}-\mu+2g( |\Phi_{0}|^2 + n_{T})\right]v_{k}({\bf r})
-g(|\Phi_{0}|^{2})u_{k}({\bf r})\nonumber\\ = \epsilon_{k}
v_{k}({\bf r})
\end{eqnarray}
In spherical symmetry, quantum numbers $n, \ell, m$ may be
attached to the quasi-particle states $k$. In what follows, we
will scale ${\bf r}$ by $r_{0} = \sqrt{\hbar/m \omega_{{\rm ho}}}$ and
energies by $\hbar \omega_{{\rm ho}}$.

Numerical methods using the discrete variable representation (DVR)
\cite{BayeHeenan} have been applied \cite{SF,BFBS} to solve the
Gross-Pitaevskii
and Bogoliubov equations. A thermal sum over discrete
quasi-particle states is supplemented at high energies by an integral
over a semiclassical representation.

\section{Asymptotic Energy Shifts}

We first consider the energy shifts of the collective and single
particle modes of harmonically trapped condensate, relative to the
energies of the harmonic oscillator modes.  We are interested in
the asymptotic behavior of the shifts as a function of the angular
momentum, $l$, and the number of radial nodes $n$ of the excited
modes. In this section we consider only zero temperature, which
implies that almost all the atoms are in the condensate.

\begin{figure}
\centering{\psfig{figure=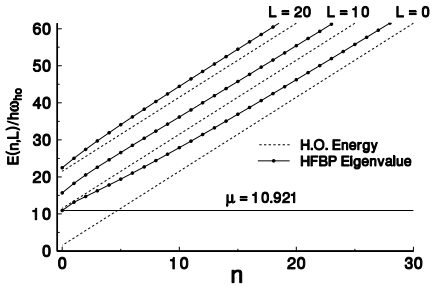,width=2.5in,angle=0}}
\caption{Hartree-Fock Bogoliubov Popov (HFBP) and harmonic
oscillator (HO) energies for $\ell$=0, 10 and 20, relative to the
trap minimum.  The chemical potential is shown as a horizontal
line. The scaled scattering length is $a/r_{0}=0.0072$ and number
of atoms $N=20,000$.} \label{Ecomp}
\end{figure}

By way of orientation, we first discuss overall trends of the energies
$\epsilon_{nl}$, or more precisely, $E_{nl} = \epsilon_{nl} + \mu$. In
Fig. \ref{Ecomp}, we have plotted numerically computed eigenvalues
$E_{nl}$, for $T=0$ and for $l$ = 0, 10 and 20 on an energy scale relative
to the bottom of the harmonic well.  On this scale, the condensate lies at
an energy equal to the chemical potential $\mu$, shifted up from the
harmonic zero point energy by the repulsive atom-atom interactions, and all
excited states lie above the condensate.  ``Bare'' harmonic oscillator
energies $E_{nl}^{0}$ are shown for comparison.  We will be interested in
how the values $E_{nl}$ approach the $E_{nl}^{0}$, but first it is relevant
to point out certain trends at low $n$.

For $\ell >  0$, centrifugal distortion pushes the excited state wavefunction
away from the condensate and thus reduces the shift for very small $n$. For
larger $n$, the inner turning point moves inwards and the overlap
with the condensate, and hence the energy shift, increases over a
certain range of $n$. Then spreading of the excited state
wavefunction to larger values of $r$ results in a decreasing shift.
These effects are fully discussed in Ref. \cite{CGSWKB} in the semiclassical
context.

A detailed and rigorous derivation of the energy shifts by WKB
methods has been presented in \cite{CGSWKB} from a consideration
of classical particle motion in the presence of the condensate.
Trajectories that do not penetrate the condensate have
excitation energies that are simply the harmonic oscillator
energies,
\begin{eqnarray}
E_{n\ell}^{0} = (2n + \ell + 3/2).
\end{eqnarray}
For trajectories that do penetrate the condensate region, there is a shift:
\begin{eqnarray} \delta_{n \ell} = E_{n \ell} - E_{n \ell}^{0} \end{eqnarray}
that is found to be \cite{CGSWKB}
\begin{eqnarray} \delta_{n\ell} = \frac{1}{3\pi}
\frac{[4\mu(2n+\ell+3/2-\mu)-(\ell - 1/2)^{2}]^{3/2}}{[2n+\ell + 3/2-\mu]^{2}}.
\label{CGSdE} \end{eqnarray}

It is possible to obtain the $n \rightarrow \infty$ limit of the
above expression by simple perturbation theory. For this purpose,
we set $n_{T}({\bf r})=0$ as appropriate for $T=0$.  We define a
$2 \times 2$ matrix Hamiltonian corresponding to the left hand
sides of the coupled equations Eq. (\ref{Bog1}):
\begin{equation}\label{bogmat}
{\bf H}_{\rm Bog}=\left(\!\!\begin{array}{cc}
    H_{\rm ho}\!-\!\mu & 0 \\0  & \!\!\!-H_{\rm ho}\!+\!\mu
    \end{array}\!\!\right)+\left(\!\!\begin{array}{cc}
    2g\left|\Phi_{0}\right|^{2} & g\left|\Phi_{0}\right|^{2}\\
    -g\left|\Phi_{0}\right|^{2}  & -2g\left|\Phi_{0}\right|^{2}
    \end{array}\!\!\right).
\end{equation}
The zeroth order eigenvector corresponding to eigenvalue $E_{nl}^{0}-\mu$
is  $\left(\psi_{nlm}({\bf r}),0\right)^{T}$, where
$\psi_{nlm}({\bf r})$ is an eigenstate of the spherical harmonic
oscillator Hamiltonian, $H_{\rm ho}$:
\begin{equation}
\psi_{nlm}(r,\theta,\phi) =
\frac{\sqrt{2n!}}{\sqrt{(n+l+\frac{1}{2})!}}L_{n}^{l+\frac{1}{2}}(r^{2})
 r^{l}e^{-\frac{1}{2}r^{2}}\frac{Y_{nlm}(\theta,\phi)}{\sqrt{4\pi}}.
\end {equation}
We have used standard notation for the spherical harmonics
$Y_{nlm}(\theta,\phi)$ and Laguerre polynomials
$L_{n}^{l+\frac{1}{2}}(r^{2})$. Considering the second matrix
operator in Eq. (\ref{bogmat}), the lowest order energy shift is:
\begin{equation}
\Delta E_{nl}=2g\langle \psi_{nlm} | \left|\Phi_{0}\right|^{2}
\psi_{nlm} \rangle. \label{DE1}
\end{equation}

An estimate of the behaviour of the shifts for large values of $n$ may be
obtained from the Thomas-Fermi approximation for the ground-state density,
\begin{equation}
 \left|\Phi_{0}\right|^{2}=\frac{2\mu-r^{2}}{2g},
\end {equation}
and the asymptotic form of the Laguerre polynomials
\begin{equation}
L_{n}^{l+\frac{1}{2}}(r^{2})
\simeq\frac{1}{\sqrt{\pi}}e^{\frac{1}{2}r^{2}}r^{-(l+1)}n^{\frac{l}{2}}
\cos\left[2\sqrt{n}r-(l+1)\frac{\pi}{2}\right].
\end{equation}
Stirling's formula gives an approximate value
\begin{eqnarray}\label{stirling1}
 \frac{n!}{(n+l+\frac{1}{2})!} \approx
\frac{1}{n^{l+\frac{1}{2}}}\frac{e^{l+\frac{1}{2}}}
{\left[\left(1+\frac{l+\frac{1}{2}}{n}\right)^{n}
 \right]^{(1+\frac{l+1}{n})}},  \label{dEnn}
\end{eqnarray}
which, for large values of $l$, can be further simplified to \begin{eqnarray}
\label{stirling2}\frac{n!}{(n+l+\frac{1}{2})!}\approx
n^{-(l+\frac{1}{2})}e^{-\frac{(l+\frac{1}{2})(l+1)}{n}}.\end{eqnarray} We
thereby obtain the asymptotic first order shift in energy
\begin{eqnarray}\label{shifteq}
\Delta E_{nl}\simeq\hspace{-1mm} \frac{
e^{-\frac{(l+\frac{1}{2})(l+1)}{n}}}{\pi\sqrt{n}}
\hspace{-1.5mm}\int^{\sqrt{2\mu}}_{0} \hspace{-7mm}dr
(2\mu\hspace{-1mm}-\hspace{-1mm}r^{2})
\hspace{-1mm}\left[1\hspace{-1mm}+(\hspace{-1mm}-1)^{l}\cos(4\sqrt{n}
r)\right] \nonumber\\ \simeq
\frac{2}{3\pi}\sqrt{\frac{(2\mu)^3}{n}}
e^{-\frac{(l+\frac{1}{2})(l+1)}{n}}\hspace{3.1cm}
\end{eqnarray}
to leading order in $n$.
In the limit $\ell \ll n \rightarrow \infty$, the shift becomes
\begin{eqnarray} \delta_{n \ell} = \frac{2}{3\pi} \sqrt{\frac{(2\mu)^{3}}{n}},
\end{eqnarray}
which agrees exactly with the asymptotic value of Eq. (\ref{CGSdE}).

\begin{figure}
\centering{\psfig{figure=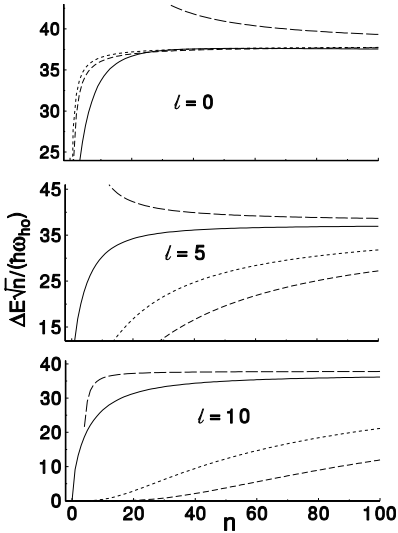,width=2.5in,angle=0}}
\caption{Energy shifts $\Delta E$ multiplied by $\sqrt{n}$ are
plotted versus $n$, the number of nodes in the radial
wavefunction.  The solid lines represent the energy shifts
obtained from numerical solutions of the Bogoliubov equations
(\ref{Bog1}). The long dashed lines above the solid lines
represent the energy shifts calculated analytically from the WKB
theory given in Eq. (10).  The medium dashed lines show
perturbation theory results of Eq. (17), while for the short
dashed lines the more exact Stirling's approximation in Eq. (15)
is inserted in Eq. (17).  In each case, the scaled scattering
length is $a/r_{0}=0.0072$ and number of atoms $N=50,000$.}
\label{Eshift}
\end{figure}

To compare the asymptotic value of the shift obtained numerically with the
value obtained from Eq. (\ref{CGSdE}), and also with values from
Eqs. (\ref{stirling1}) through (\ref{shifteq}),
we plot these respective values multiplied by
$\sqrt{n}$ over a range of values of $n$ for $l = 0$, 5 and
10 in Fig. \ref{Eshift} (for $N=50,000, a_{sc}/r_{0} = 0.0073$).
For $\ell = 0$ and also $\ell = 1$, the simple perturbation
theoretic results happen to be closer to the numerical results
than Eq. (\ref{CGSdE}), but for all high values of $\ell$ and
increasingly so, the WKB result is much closer than the
perturbation theoretic result with asymptotic Laguerre functions.
However, the WKB result does tend to approach the asymptote more
rapidly than the numerical results.  It should also be noted that
the use of the more accurate expression, Eq. (\ref{stirling1}) for
Stirling's formula produces an improvement over the usual approximation.

We have also calculated first order energy shifts, Eq. (\ref{DE1})
with exact harmonic oscillator wavefunctions rather than the asymptotic
forms given above.  The results are extremely close to the numerical
results obtained with quasi-particle wavefunctions, from Eq. (\ref{Bog1}).

\begin{figure}
\centering{\psfig{figure=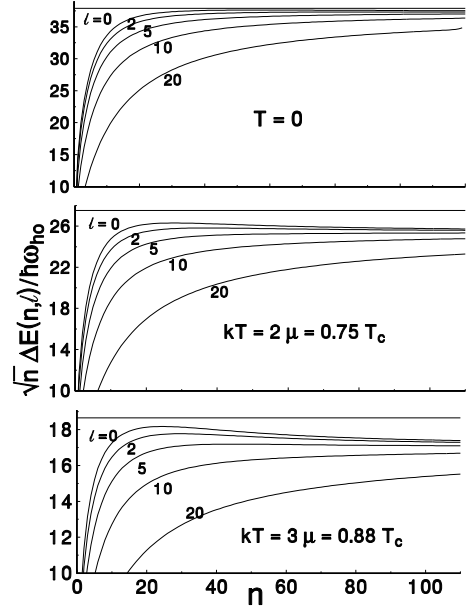,width=3.2in,angle=0}}
\caption{Energy shifts $\Delta E$=HFBP($n,\ell$) - HO($n,\ell$)
multiplied by the $\sqrt{n}$ are plotted for different values of
$\ell$, showing that the asymptotic shift is independent of
$\ell$, and approximately equal to $2 \sqrt{(2 \mu)^{3}}/3\pi$.
The scaled scattering length is $a/r_{0}=0.0072$ and number of
atoms $N=50,000$.}\label{Edifl}
\end{figure}

The fact that the asymptotic shift is independent of $\ell$ is
demonstrated in Fig. \nolinebreak \ref{Edifl} where we have plotted
the energy shifts multiplied by $\sqrt{n}$ for $l$ = 0, 5, 10 and 20.
These shifts can be contrasted with the behaviour
of Rydberg states of closed shell atoms, for which the asymptotic shift
is $R \delta_{\ell}/n^{3}$, where the dipole polarizability is
$\delta^{pol}_{\ell}\simeq 3\alpha_d/4 \ell^5$, with $\alpha_d$ the dipolar
polarizability. Thus for atomic Rydberg states, the asymptotic shift
exhibits a sharp $\ell$ dependence that is lacking here. One other contrast
with the theory of atomic Rydberg states is that the ion core can be considered
effectively localized at the origin, while for excitations of a Bose
condensate, a spatial integral, as in Eq. (\ref{shifteq}), is essential.

In cylindrical symmetry, WKB wavefunctions exhibit chaotic
behaviour \cite{Fliesser1}, and the analysis cannot be carried out
as for spherical symmetry. The asymptotic shift may be obtained
from perturbation theory in a manner analogous to the above
discussion, and results in the following expression:
\begin{equation}
\Delta E_{n_{\rho},n_{z},m} =\frac{\mu^{2}e^{-
\frac{m^2}{n_\rho}}}{8\pi^{\frac{1}{2}}
\sqrt{\omega_{\rho}n_{\rho}\omega_{z}n_{z}}}
\end{equation}
where $\omega_\rho$ and $\omega_z$ are trap frequencies, and $m$ is an exact
quantum number. However, $n_\rho$ and $n_z$ are exact only when there is no
condensate present.

\section{Landau and Beliaev Damping of Excitations in a Trapped Bose
Gas} \label{dmping}

We now consider the damping of these collective and single particle
excitations.  An
estimate of the width of the modes is essential to establish the
feasibility of resolving high-lying excitations of a
Bose condensate.  The primary source of
damping is the interaction of the collective excitation of the
condensate with the noncondensate or thermal part.  For high-lying
excitations, the possibilities for both Landau and Beliaev decay
increase with temperature, so it is interesting to obtain these
rates explicitly.

A detailed exposition of the theory of Landau and Beliaev damping of
quasi-particle excitations for Bose condensates has been presented by
Giorgini \cite{G1}. The expressions for Landau damping were in agreement
with expressions given earlier by Pitaevskii and Stringari \cite{PS}, but
the derivation was considerably more detailed.  Giorgini's derivation is
applicable to Bose atoms in harmonic trap potentials as well as to
homogeneous Bose gases. Our discussion takes another point of view
that results in expressions with a Lorentz profile. Consequently,
all interacting levels within the linewidth of a resonant mode
contribute to damping by Landau or Beliaev mechanisms.
We note again that references \cite{Fedichev1} and \cite{Reidl2} also
present theories of Landau damping for trapped Bose atoms.

\subsection{Damping from Perturbation Theory}\label{WWSec}

We first review some of the methods of time-dependent perturbation
theory \cite{Fetter,Heitler} that will be used in our approach to the
calculation of the widths. It will be convenient to work in the
interaction picture, in which the
time dependence of the operators and the states is given by
\begin{eqnarray}
\hat{O}(t)=e^{i\hat{H}_{0}t/\hbar}\hat{O}_{S}e^{-i\hat{H}_{0}t/\hbar}
\end{eqnarray}
\begin{eqnarray}|\Psi(t)\rangle=e^{i\hat{H}_{0}t/\hbar}|\Psi_{S}(t)\rangle,
\end{eqnarray}
where $\hat{O}_{S}$ signifies a time-independent operator in the
Schr\"{o}dinger picture. The Hamiltonian is $\hat{H}=\hat{H}_{0}+\hat{V}$.
The differential equations that determine the time
development are \begin{eqnarray}\label{timdep}i\hbar\frac{\partial}{\partial t}
|\Psi(t)\rangle=\hat{V}(t)|\Psi(t)\rangle\nonumber\\
i\hbar\frac{\partial}{\partial t}\hat{O}(t)
=[\hat{O}(t),\hat{H}_{0}].\hspace{1mm}
\end{eqnarray}

The time development of the state from time $t=0$ to a later
time $t$ can be effected by means of a time-evolution operator
$\hat{U}(t,0)$ acting on the initial state \begin{eqnarray}
|\Psi(t)\rangle=\hat{U}(t,0)|\Psi(0)\rangle.\end{eqnarray} From
Eq. (\ref{timdep}) it follows that the operator satisfies the
differential equation \begin{eqnarray} i\hbar\frac{\partial}{\partial
t}\hat{U}(t,0)=\hat{V}(t)\hat{U}(t,0)\end{eqnarray} which  can be
integrated with the initial condition
$\hat{U}(0,0)=1$ to give \begin{eqnarray} \label{Uevolve}
\hat{U}(t,0)=1-\frac{i}{\hbar}\int^{t}_{0}dt'
\hat{V}(t')\hat{U}(t',0).\end{eqnarray}

The final state can be expanded in a basis of stationary
eigenstates of $\hat{H}_0$ \begin{eqnarray}
|\Psi(t)\rangle=\sum_{n}a_{n}(t)|\Psi_{n}\rangle,\end{eqnarray}
with coefficients \begin{eqnarray} \label{andef}a_{n}(t)=\langle
\Psi_{n}|\hat{U}(t,0)|\Psi(0)\rangle
\hspace{4cm}\nonumber\\=\langle \Psi_{n}
|\Psi(0)\rangle-\hspace{-1mm}\sum_{m}\frac{i}{\hbar}\int^{t}_{0}
\hspace{-2mm}dt'\langle \Psi_{n}| V(t')|\Psi_{m}\rangle
a_{m}(t').\end{eqnarray}

At this point we make the simplifying assumption that the initial
state is an eigenstate of $\hat{H}_{0}$,
$|\Psi(0)\rangle=|\Psi_{i}\rangle$. Moreover, if the time
scales are short enough and there are a large number of possible
final eigenstates, it can be assumed to a very good approximation
that $a_{i}(t)\gg a_{n\neq i}(t)$, so that \begin{eqnarray}\label{aneq}
a_{n\neq i}(t) \simeq-\frac{i}{\hbar}\int^{t}_{0}dt'\langle
\Psi_{n}|\hat{V}(t')|\Psi_{i}\rangle a_{i}(t'). \end{eqnarray}

If the time scales are sufficiently short that changes in the
initial state can be ignored, we can set
$a_{i}(t)\approx 1$, and Eq. (\ref{aneq}) can be solved for each $n \neq i$
to obtain:
\begin{eqnarray} |a_{n}(t)|^{2}=2|\langle
\Psi_{n}|V_{S}|\Psi_{i}\rangle|^{2}\left[
\frac{1-\cos\{(\omega_{n}-\omega_{i})t\}}{[\hbar(\omega_{n}-\omega_{i})]^{2}}
\right].\end{eqnarray}
We have reverted to the Schrodinger picture in order to make the
time-dependence explicit. $E_{n}=\hbar\omega_{n}$ is the energy associated
with the $n$th eigenstate of the unperturbed Hamiltonian $H_{0}$.
Then the usual procedure is to assume that $t\gg \hbar/(E_{n}-E_{i})$, so that
even though $t$ is still small compared with the inverse decay rate, one is
justified in writing the  quantity in brackets as a delta function. In this way,
one obtains Fermi's Golden Rule:
\begin{eqnarray} \Gamma_{n} = \frac{d}{dt}|a_{n}(t)|^{2}
=\frac{2\pi}{\hbar}|\langle
\Psi_{n}|V_{S}|\Psi_{i}\rangle|^{2}\delta({E_{n}-E_{i}}). \label{FermiGR}
\end{eqnarray}
The sum of the partial widths, $\sum_{n} \Gamma_{n}= \Gamma$, gives the total
decay rate at $t=0$.

This expression serves for the evaluation of transition probabilities
in a wide range of physical problems. However, for application to decay
rates, it requires that there be an intermediate range of times that are large
compared with the oscillation periods but small compared with $\Gamma^{-1}$.
For a typical radiative decay process, for example a D line of sodium with
$\Gamma/\omega \approx 2 \times 10^{-8}$, this is no problem.
For the resonance excitations of Bose condensates reported in
Ref. \cite{Jin2}, however, $\Gamma/\omega \approx 0.01$ to 0.1, which makes
the $\delta$ function in Eq. (\ref{FermiGR}) problematical. Hence it is
appropriate to introduce the method of Weisskopf and Wigner \cite{Heitler,WW},
which removes the upper limit on the range of $t$ by dropping the assumption
that $a_{i}(t) \approx 1$. This is done by introducing an exponential decay
of the initial state:
\begin{eqnarray} \label{expdep}a_{i}(t)=e^{-\frac{1}{2}\Gamma t}.
\end{eqnarray}
Eq. (\ref{aneq}) then becomes
\begin{eqnarray}\label{aneq1}
a_{n}(t)\simeq-\frac{i}{\hbar}\int^{t}_{0}dt'\langle
\Psi_{n}|\hat{V}_{S}|\Psi_{i}\rangle
e^{i(\omega_{n}-\omega_{i})t'} e^{-\frac{1}{2}\Gamma t'}
\nonumber\\=\langle \Psi_{n}|\hat{V}_{S}|\Psi_{i}\rangle
\frac{1-e^{i(\omega_{n}-\omega_{i})t} e^{-\frac{1}{2}\Gamma
t}}{\hbar\{(\omega_{n}-\omega_{i})+i\Gamma/2\}}. \hspace{.8cm}
\end{eqnarray}
Conservation of probability for a complete set of
states for a closed system \begin{eqnarray}\label{unitarity}
1-|a_{i}(t)|^2=\sum_{m\neq i}|a_{m}(t)|^2
\end{eqnarray} then gives \begin{eqnarray}
1-e^{-\Gamma t}= \sum_{m\neq i}\frac{|\langle \Psi_{m}|
\hat{V_{S}}|\Psi_{i}\rangle|^{2}}
{\hbar^{2}\{(\omega_{m}-\omega_{i})^{2}+\left(\Gamma/2\right)^{2}\}}
\hspace{1.5cm}\nonumber\\ \times\left[1+e^{-\Gamma t}-2e^{-\Gamma
t/2} \cos\{(\omega_{m}\!-\!\omega_{i})t\}\right]. \end{eqnarray}

As it stands, this expression does not have any assumptions about
the time-scales involved.  We will take the long term limit,
$t\rightarrow \infty$, in order to remove the explicit dependence
on time \begin{eqnarray}\label{maineq} \Gamma/2=\sum_{m\neq
i}\frac{|\langle\Psi_{m}|\hat{V_{S}}|\Psi_{i}\rangle|^{2}}{\hbar^2}
\frac{\Gamma/2}
{(\omega_{m}-\omega_{i})^{2}+\left(\Gamma/2\right)^{2}}\end{eqnarray}

We have multiplied both sides by $\Gamma/2$ to obtain an implicit
equation for the width that can be solved iteratively with a
knowledge of the stationary states and their energies. This is the general
form we will use to obtain decay rates for excitations in a Bose gas.

\subsection{The Damping Interaction}

We now consider the operators and the basis states that are
relevant for a Bose-condensed system. Since we will work within
the grand canonical ensemble, $H_{0}$ in the previous section will
be replaced by $K_{0}$, which is the part of the grand canonical
Hamiltonian in Eq. (\ref{Kdef}) that is diagonalized by the
Bogoliubov transformation, Eq. (\ref{Bog}). The eigenfunctions are
the quasi-particle basis states as given in equations
Eqs. (\ref{GPE})and (\ref{Bog1}). The quasi-particles themselves
are considered to be in a thermal distribution determined by
$K_{0}$, so that the relevant statistical operator is
\begin{eqnarray}
\frac{e^{-\beta\hat{K}_{0}}}{{\rm Tr}
\hspace{1mm}e^{-\beta\hat{K}_{0}}}. \end{eqnarray}
Eigenstates of $K_{0}$ may
be written in a Fock representation of quasi-particle occupation
numbers $ |n_{1},n_{2},\cdots\rangle$, and satisfy
\begin{eqnarray}
\hat{K}_{0}|n_{1},n_{2},\cdots\rangle=\left[\sum_{k} \epsilon_{k}
n_{k}\right]|n_{1},n_{2},\cdots\rangle\end{eqnarray} where
$\epsilon_k=\hbar\omega_{k}$ is the energy and $n_{k}$ is the occupation
number of the $k$th quasi-particle mode. The creation and
destruction operators for the quasi-particle modes are defined by
\begin{eqnarray} \hat{\alpha}_{k}|n_{1},\cdots,n_{k},\cdots\rangle=
\sqrt{n_{k}}|n_{1},\cdots,n_{k}-1,\cdots\rangle\hspace{7mm}\nonumber\\
\hat{\alpha}^{\dagger}_{k}|n_{1},\cdots,n_{k},\cdots\rangle=
\sqrt{n_{k}+1}|n_{1},\cdots,n_{k}+1,\cdots\rangle.\end{eqnarray}
They obey the standard Bose commutator relations and their
time-dependence in the interaction picture follows from Eq.
(\ref{timdep}) \begin{eqnarray}
\hat{\alpha}_{k}(t)=\hat{\alpha}_{k}e^{-i\omega_{k}t}\hspace{1cm}
\hat{\alpha}^{\dagger}_{k}(t)=\hat{\alpha}_{k}e^{i\omega_{k}t}.\end{eqnarray}
In this representation, the matrix elements of the statistical
operator are \begin{eqnarray} \frac{\left\langle
n_{1},n_{2},\cdots\right|e^{-\beta\hat{K}_{0}}
\left|n_{1},n_{2},\cdots\right\rangle}{\sum_{\{m_{1}\}}
{\left\langle m_{1},m_{2},\cdots\right|e^{-\beta\hat{K}_{0}}
\left|m_{1},m_{2},\cdots\right\rangle}}
\nonumber\\=\frac{\prod_{k} e^{-\beta
\epsilon_{k}n_{k}}}{\prod_{k} \sum_{m_k}e^{-\beta
\epsilon_{k}m_{k}}}.\hspace{3.2cm}\end{eqnarray}

We next define the interaction that causes damping. We start with
the interaction term in Eq. (1), separate the condensate
wavefunction as done in Eq. (\ref{Bogdec}), introduce the small
amplitude fluctuation of Eq. (\ref{flucsep}) and linearize with
respect to the fluctuation to get \begin{eqnarray} \frac{g}{2} \!
\int \! d{\bf r} \hat{\psi}^{\dagger}\hat{\psi}^{\dagger}
\hat{\psi}\hat{\psi} \approx g\!\! \int \!\! d{\bf r} \left\{
\Phi_{0}\Phi_{0}\Phi_{0}(\delta\hat{\Phi}+\delta\hat{\Phi}^{\dagger})
\hspace{1cm}\right.\nonumber\\ \left.
+\Phi_{0}\Phi_{0}[(2\delta\hat{\Phi}+\delta\hat{\Phi}^{\dagger})
\hat{\phi}^{\dag}+ (\delta\hat{\Phi}+2\delta\hat{\Phi}^{\dagger})
\hat{\phi}]\right.\nonumber\\ \left.
+\Phi_{0}[\delta\hat{\Phi}^{\dagger}\phi\phi+
2(\delta\hat{\Phi}+\delta\hat{\Phi}^{\dagger})\hat{\phi}^{\dag}\hat{\phi}
+\delta\hat{\Phi}^{\dagger}\hat{\phi}^{\dag}\hat{\phi}^{\dag}]
\right.\nonumber\\ \left. +(\delta\hat{\Phi}
+\delta\hat{\Phi}^{\dagger})\hat{\phi}^{\dag}
\phi\phi\right\}\end{eqnarray} Here, and in subsequent
expressions, for the sake of brevity, the spatial and temporal
dependencies are not shown explicitly where they are obvious. In
this paper, we assume that the condensate fluctuation mode is
identical with one of the quasi-particle modes Eq. (\ref{Bog}),
which we designate by the subscript {\it`os'} :
\begin{eqnarray}\label{delPhi} \delta\hat{\Phi}({\bf r},t)=u_{os}({\bf
r})\hat{\alpha}_{os}(t)\ +\ v^{*}_{os}({\bf
r})\hat{\alpha}^{\dagger}_{os}(t).\end{eqnarray} We ignore the dynamics of
the noncondensate, {\it i.e.} the noncondensate quasi-particles are
assumed to be in static thermal equilibrium, with a population
given by the Bose distribution function.  As in \cite{GP}, we
regard the effect of the oscillation as an increase of the
quasi-particle population of the mode labeled {\it`os'} from its
value, $f^{0}_{os}$ in a thermal distribution to a macroscopic
value, which we call $n_{os}$. Furthermore, we assume that the
amplitude of the oscillation is small enough that the thermal
equilibrium of the noncondensate particles is not affected.

Before making the quasi-particle substitutions, we note that the
terms that are zeroth order in the noncondensate operators
cannot contribute to the decay of the fluctuation due to
conservation of energy.  The first order terms will also be
constrained by conservation of energy so that each quasi-particle
removed from the condensate fluctuation will be compensated by an
identical quasi-particle added to the noncondensate; this
essentially leaves the number of quasi-particles in the condensate
fluctuation unchanged. Thus the lowest order contribution to the
decay of the condensate fluctuation comes from the terms bilinear
in the noncondensate operators.  Hence, the damping interaction
is defined by \begin{eqnarray} \hat{V}(t)=g\!\!\int\!\! d{\bf
r}\,\Phi_{0}[\delta\hat{\Phi}^{\dagger}\hat{\phi}\hat{\phi}
\!+\!2(\delta\hat{\Phi}\!+\!\delta\hat{\Phi}^{\dagger})
\hat{\phi}^{\dag}\hat{\phi}
\!+\!\delta\hat{\Phi}\hat{\phi}^{\dag}\hat{\phi}^{\dag}]\end{eqnarray}

We transform to quasi-particles and retain all the terms that do
not violate conservation of energy. These are grouped into terms (labeled
$\hat{V}^{+}(t)$) that create a quasi-particle in the condensate
fluctuation and those that annihilate one (labeled $\hat{V}^{-}(t))$:
\begin{eqnarray} \label{pertpot}\hat{V}(t)=\hat{V}^{+}(t)+\hat{V}^{-}(t).
\end{eqnarray}

These terms include two processes: (i) Landau decay in which a
quasi-particle from the condensate fluctuation interacts with a
noncondensate quasi-particle to create a higher energy
quasi-particle and (ii) Beliaev decay in which a condensate fluctuation
quasi-particle decays into two lower energy quasi-particles.
The two processes are schematically represented below
\begin{eqnarray}
E_{os}+E_{j}\leftrightarrow E_{i} \hspace{1cm} \rm{Landau} \nonumber \\
E_{os}\leftrightarrow E_{i}+E_{j}. \hspace{1cm}
\rm{Beliaev} \label{LB} \end{eqnarray}
$\hat{V}^{-}(t)$ correspond to these processes going from left to
right while those included under $\hat{V}^{+}(t)$ proceed in the
reverse direction.

At this point, we make the
time-dependence of the quasi-particle operators explicit. Also,
since we are considering a stationary condensate the
quasi-particle wave functions $u_{k}$ and $v_{k}$ are taken to be
real. After some algebra and rearrangements of the indices under the
summation, we obtain
\begin{eqnarray}\label{pertpot1}\hat{V}^{+}(t)\!=\!\!\sum_{i,j}\!\!\left[L_{ij}
e^{i\omega^{-}_{ij} t}\hat{\alpha}_{os}^{\dagger}
\hat{\alpha}_{j}^{\dagger}\hat{\alpha}_{i}
+B_{ij}e^{i\omega^{+}_{ij}t}
\hat{\alpha}^{\dagger}_{os}\hat{\alpha}_{j}\hat{\alpha}_{i}\right]\nonumber\\
\hat{V}^{-}(t)\!=\!\!\sum_{i,j}\!\!\left[L_{ij}e^{\!\!-i\omega^{-}_{ij}
t}\hat{\alpha}_{os} \hat{\alpha}_{j}\hat{\alpha}^{\dagger}_{i}
+B_{ij}e^{\!\!-i\omega^{+}_{ij}t}
\hat{\alpha}_{os}\hat{\alpha}^{\dagger}_{j}\hat{\alpha}^{\dagger}_{i}
\right]\hspace{-3mm}
\end{eqnarray}
with $\omega^{\pm}_{ij}=\omega_{os}-(\omega_{i}\pm \omega_{j})$
and \begin{eqnarray}\label{pertpot2} L_{ij}=2g \int d{\bf r}\
\Phi_{0} \left[u_{os} (u^{}_{i}v_{j}+u^{}_{i}u_{j}
+v^{}_{i}v_{j})+\hspace{.5cm}\right. \nonumber\\ \left. v_{os}
(u^{}_{i}v_{j}+u^{}_{i}u_{j} +v^{}_{i}v_{j}) \right]\nonumber
\\B_{ij}=g\int d{\bf r}\ \Phi_{0} \left[u_{os}(u^{}_{i}u^{}_{j}
+v^{}_{i}u^{}_{j}+v^{}_{j}u^{}_{i})
\hspace{.9cm}\right.\nonumber\\
\left.+v_{os}(v^{}_{j}u^{}_{i}+v^{}_{i}v^{}_{j}
+v^{}_{i}u^{}_{j})\right].\end{eqnarray}

\subsection{\bf Calculation of the Width}

The expressions for the interaction derived in the
previous section can now be applied to the methods developed in Sec.
IV.A to  obtain the decay rate. Since we consider finite
temperatures, the system has to be described by a density matrix.
We incorporate our considerations from the previous section into
the definition of the density matrix. The condensate oscillation
is treated as an increase in population of a particular mode,
labeled with subscript `{\it os}', to a macroscopic value
$n_{os}\gg 1$. At the same time the enhanced population $n_{os}$
is considered sufficiently small to leave the remaining modes well
approximated by a thermal distribution.  Thus we define the initial
density matrix to be
 \begin{eqnarray}\label{denmat}
\rho(0)=\sum_{\stackrel{\{n_{k}\}}{k\neq os}}
W^{\{n_{k}\}}|n_{1},n_{2}..n_{os}..\rangle \langle
n_{1},n_{2}..n_{os}..|.\end{eqnarray}
It is characterized by a well-defined
and fixed value of $n_{os}$, for the condensate oscillation mode,
whereas the other modes are thermally distributed over all
possible values. This is reflected in the statistical weight
\begin{eqnarray} W^{\{n_{k}\}}=\frac{\prod_{k\neq os
}e^{-\beta \epsilon_{k}n_{k}}}{\prod_{l\neq os}
\sum_{m_l}e^{-\beta \epsilon_{l}m_{l}}}.\end{eqnarray}
The time evolution of
the diagonal matrix elements of the density matrix is given by
\begin{eqnarray}\label{denmatevol} \langle m_{1},m_{2},..|\rho(t)|
m_{1},m_{2},..\rangle\hspace{3cm}\nonumber\\=\langle
m_{1},m_{2},..|\hat{U}(t,0)\rho(0)\hat{U}^{\dagger}(0,t)|m_{1},m_{2},..\rangle
\hspace{8mm}\nonumber \\ =\sum_{\stackrel{\{n_{k}\}}{k\neq os}}
W^{\{n_{k}\}}|\langle
m_{1},m_{2},..|\hat{U}(t,0)|n_{1},n_{2}..n_{os}..\rangle|^{2},\end{eqnarray}
where, $\hat{U}(t,0)$ is the time evolution operator defined
earlier. We will use it in the linear approximation
\begin{eqnarray} \hat{U}(t,0)\simeq {\mathbf
1}-\frac{i}{\hbar}\int_{0}^{t}dt'\hat{V}(t').\end{eqnarray} The
states $|m_{1},m_{2},\cdots\rangle$  that evolve from the initial
state can have arbitrary occupation numbers for {\it all} modes.
For damping of the condensate oscillation mode, however, only
states with $m_{os}\neq n_{os}$ are relevant, for which case
\begin{eqnarray}\label{diagel} \langle m_{1},m_{2},..|\rho(t)|
m_{1},m_{2},..\rangle \hspace{3.3cm}\nonumber\\ \simeq\!\!
\sum_{\stackrel{\{n_{k}\}}{k\neq os}}\!\!
\frac{W^{\{n_{k}\}}}{\hbar^{2}}\!\left|\int^{t}_{0}\!\!\!\!dt'
\langle m_{1},m_{2},..| \hat{V}(t')|n_{1},n_{2}..n_{os}..\rangle
\right|^{2}\end{eqnarray}

Now we introduce an {\it Ansatz} similar to the Weisskopf-Wigner
hypothesis that the number of particles in the condensate
oscillation mode decays exponentially, {\it i.e.} \begin{eqnarray} \label{N0t}
n_{os}(t)=n_{os}e^{-\Gamma t}\end{eqnarray}
This {it Ansatz} and the interaction derived in
Eqs.(\ref{pertpot})-(\ref{pertpot2}) allow us to evaluate the matrix
elements in Eq. (\ref{diagel})
\begin{eqnarray}
\label{tranamp1}\left|\int^{t}_{0}\!\!\!\!dt'\langle
m_{1},m_{2},\cdots|\hat{V}^{+}(t')|n_{1},n_{2}..n_{os}..\rangle\right|^{2}
=\hspace{1cm}\nonumber\\L^{2}_{ij}\left|\int^{t}_{0}\!\!\!\!dt'e^{(i\omega^{-}_{ij}
-\frac{1}{2}\Gamma)t'}\right|^{2}\!\!\!\!(n_{os}+1)(n_{j}+1)
n_{i}\!\left|\!\!\begin{array}{c}
  _{m_{os}=n_{os}+1}\\
  _{m_{j}=n_{j}+1} \\
  _{m_{i}=n_{i}-1}
\end{array}\right.
\nonumber \\ +
B^{2}_{ij}\left|\int^{t}_{0}\!\!\!\!dt'e^{(i\omega^{+}_{ij}-\frac{1}{2}\Gamma)
t'}\right|^{2}\!\!\!\!(n_{os}+1)n_{j}n_{i}\!\left|\!\!\begin{array}{c}
  _{m_{os}=n_{os}+1}\\
  _{m_{j}=n_{j}-1} \\
  _{m_{i}=n_{i}-1}
\end{array}\right.\end{eqnarray}

 \begin{eqnarray}
\label{tranamp2}\left|\int^{t}_{0}\!\!\!\!dt'\langle
m_{1},m_{2},\cdots|\hat{V}^{-}(t')|n_{1},n_{2}..n_{os}..\rangle\right|^{2}
=\hspace{1cm}\nonumber\\L^{2}_{ij}\left|\int^{t}_{0}\!\!\!\!dt'e^{(i\omega^{-}_{ij}
-\frac{1}{2}\Gamma) t'}\right|^{2}\!\!\!
n_{os}n_{j}(n_{i}+1)\!\left|\!\!\begin{array}{c}
  _{m_{os}=n_{os}-1}\\
  _{m_{j}=n_{j}-1} \\
  _{m_{i}=n_{i}+1}
\end{array}\right.
\hspace{1cm}\nonumber \\ +
B^{2}_{ij}\left|\int^{t}_{0}\!\!\!\!dt'e^{(i\omega^{+}_{ij}-\frac{1}{2}\Gamma)
t'}\right|^{2}\!\!\!\!n_{os}(n_{j}+1)(n_{i}+1)\!\left|\!\!\begin{array}{c}
  _{m_{os}=n_{os}-1}\\
  _{m_{j}=n_{j}+1} \\
  _{m_{i}=n_{i}+1}
\end{array}\right.\end{eqnarray}
It is obvious that for any given set of occupation numbers
$\{m_{1},m_{2}..\}$, only one of the above terms will differ from
zero.

We add up all the non-zero elements that arise from
Eq. (\ref{tranamp2}) considering all possible values of
$m_{1},m_{2},\cdots$, and subtract from them all the non-zero
elements that arise from Eq. (\ref{tranamp1}).  This gives us the
probability for the number of particles in the condensate
oscillation mode decreasing by one after time `{\it t}' when we
start with $n_{os}$ particles in that mode. We use the macroscopic
nature of the oscillation mode, which means $n_{os}$ is large
compared to $1$, to set $n_{os}\simeq n_{os}+1$ and obtain
\begin{eqnarray}
\label{Wn} P(n_{os},t;-1)\hspace{6cm}\nonumber\\=
\sum_{\stackrel{\{n_{k}\}}{k\neq os}} \!W^{\{n_{k}\}}
\frac{n_{os}}{\hbar^2}\!\! \sum_{i,j}\!\left[L_{ij}^{2}
\left|\int^{t}_{0}\!\!\!dt'e^{(i\omega^{-}_{ij}-\frac{1}{2}\Gamma)
t}\right|^{2}\!\!\!\!\left(n_{j}-n_{i}\right) \right.\nonumber\\
\left.
+B_{ij}^{2}\left|\int^{t}_{0}\!\!\!dt'e^{(i\omega^{+}_{ij}-\frac{1}{2}\Gamma)t}
\right|^{2}\left(1+n_{j}+n_{i}\right)\right].\end{eqnarray} The
sum over $\{n_{k}\}$ with the statistical weights essentially
reproduces a thermal distribution of the modes of other than the
condensate oscillation mode. In order to evaluate this, we
consider a generic term, \begin{eqnarray} \label{fj}
\sum_{n_{1},n_{2},\cdots}\frac{\prod_{k}e^{-\beta
\epsilon_{k}n_{k}}}{\prod_{l} \sum_{m_l}e^{-\beta
\epsilon_{l}m_{l}}} n_{j} \hspace{3.5cm}\nonumber\\=
\frac{\prod_{k=1}^{j-1}\hspace{-1mm}\prod_{k=j+1}^{\infty}\hspace{-1mm}
\left[\sum_{n_k}\hspace{-1mm}e^{-\beta \epsilon_{k}n_{k}}\right]
\sum_{n_j}\hspace{-1mm}n_{j}e^{-\beta \epsilon_{j}n_{j}}}
{\prod_{l}\sum_{m_l}e^{-\beta \epsilon_{l}m_{l}}}\nonumber\\=
\frac{ \sum_{n_j}n_{j}e^{-\beta \epsilon_{j}n_{j}}}
{\sum_{n_j}e^{-\beta \epsilon_{j}n_{j}}}=\frac{1} {e^{\beta
\epsilon_{j}}-1}\hspace{2.3cm}\end{eqnarray} Thus the thermal sum
has the effect of replacing \begin{eqnarray} n\rightarrow
f_{j}=[e^{\beta \epsilon_{j}}-1]^{-1}.\end{eqnarray}

We divide by $n_{os}$ to obtain the probability of the change per
initial quasi-particle:
\begin{eqnarray}\label{transprob1}
P(1,t;-1)\hspace{5.3cm}\nonumber\\=\frac{1}{\hbar^2}\sum_{i,j}\left[L_{ij}^{2}
\left|\int^{t}_{0}dt'e^{(i\omega^{-}_{ij}-\frac{1}{2}\Gamma)
t}\right|^{2} \left(f_{j}-f_{i}\right) \hspace{.5cm}\right.\\
\nonumber\left.
+B_{ij}^{2}\left|\int^{t}_{0}dt'e^{(i\omega^{+}_{ij}-\frac{1}{2}\Gamma)t}
\right|^{2}\left(1+f_{j}+f_{i}\right)\right].\end{eqnarray}

The time integrals can be treated in the same way as
in our general discussion in Section \ref{WWSec}. In the long time
limit i.e. $t\rightarrow \infty$, these integrals tend to \begin{eqnarray}
\left|\int^{t}_{0}dt'
e^{(i\omega^{\pm}_{ij}-\frac{1}{2}\Gamma)t'}\right|^{2}\hspace{-2mm}\rightarrow
\frac{1}{(\omega^{\pm}_{ij})^{2} +(\Gamma/2)^{2}}.
\end{eqnarray}
As for Eq. (\ref{maineq}), in the limit $t \rightarrow \infty$, the
condensate oscillation has damped out completely so that $P(1,t;-1) \rightarrow 1$, and we can obtain an implicit equation for $\Gamma$. At this point, we
introduce $\gamma = \Gamma/2$, which corresponds to the width parameter
$\gamma$ used in references \cite{PS,G1,GP}. Thus
\begin{eqnarray}\label{transprob3}\gamma=\frac{1}{\hbar^2}\sum_{i,j}
\left[L_{ij}^{2}\frac{\gamma}{(\omega^{-}_{ij})^{2}
+\gamma^{2}}\left(f_{j}-f_{i}\right)\hspace{1cm}
\hspace{1cm}\right.\nonumber\\ \left.
+B_{ij}^{2}\frac{\gamma}{(\omega^{+}_{ij})^{2}
+\gamma^{2}}\left(1+f_{j}+f_{i}\right)\right],\end{eqnarray} where
we have multiplied both sides by $\gamma=\Gamma/2$ as in Eq.
(\ref{maineq}). In the numerical work, we will use this expression
iteratively to calculate the width.

To deduce an approximate relation for the dependence of $\gamma$ on the
coupling parameter, $g$, and hence on $a$, the scattering length, we consider
just those $(i,j)$ values for which $\omega_{ij}^{\pm} < \gamma$. For
this subset of interactions, which include the largest contributions
to the width, we have
\begin{eqnarray} \label{gamapp}
\gamma^{2} \approx \frac{1}{\hbar^2}\sum_{i,j}
\left[L_{ij}^{2} \left(f_{j}-f_{i}\right)
+B_{ij}^{2} \left(1+f_{j}+f_{i}\right)\right].\end{eqnarray}
Since $L_{ij}$ and $B_{ij}$ from Eq. (\ref{pertpot2}) are each linearly
dependent on $g$, we conclude that $\gamma$ is approximately linearly dependent
on $g$ and $a$.

\subsection{Comparisons and Caveats}

Previous expressions for the damping rate given in Refs. \cite{PS,G1,GP}
employed a delta function, as in Fermi's Golden Rule, Eq. (\ref{FermiGR}).
For example, for Landau damping, Eq. (41) of Ref. \cite{G1} is
\begin{eqnarray}
\gamma_{L} = 4 \pi g^{2} \sum_{ij} |A_{ij}|^{2} (f_{i}^{0} - f_{j}^{0})
\delta(\hbar \omega_{0} + \epsilon_{i} - \epsilon_{j}). \label{G41}
\end{eqnarray}
For a manifold of discrete states, this delta function would never be
precisely satisfied.  By assuming exponential decay, we have obtained
a width expression involving a Lorentz function with a finite width,
that can be solved iteratively for $\gamma$. Our discussion in Sec. IV.A
also makes clear that the delta function is difficult to justify when the ratio
$\gamma/\omega$ approaches values of 0.01 to 0.1 as is the case for
excitations of Bose condensates. However, one can formally obtain a
similar expression from Eq. (\ref{G41}) by replacing $\omega_{0}$,
the unperturbed resonance frequency, with the perturbed $\omega$, which
is then assumed to be distributed over a range of values characterized
by a Lorentz profile with a width $\gamma$.

In the above derivation, we have explicitly calculated the decay of
diagonal density matrix elements. However, if we consider $\hat{\alpha}_{os}$
as a $c$-number, $\hat{\alpha}_{os} = \hat{\alpha}_{os}^{\dagger} =
\sqrt{N_{os}}$, and evaluate the time evolution of the squared amplitude of the
condensate wavefunction \cite{Edwards}
\begin{eqnarray} \label{super}
|\Phi({\bf r},t)|^{2}= |\Phi_{0}({\bf r})+\delta\Phi({\bf
r},t)|^2= \hspace{2.7cm}\nonumber\\|\Phi_{0}({\bf
r})|^{2}\hspace{-1mm}+\hspace{-1mm}2\Phi_{0}({\bf r})\sqrt{N_{os}}
[u_{os}({\bf r})\hspace{-1mm}+\hspace{-1mm}v_{os}({\bf r})]
\cos(\omega_{os} t) e^{-\gamma t} \nonumber\\ + N_{os}[u_{os}({\bf
r})^{2}  + v_{os}({\bf r})^{2}
\hspace{3cm}\nonumber\\+2u_{os}({\bf r})v_{os}({\bf r})
\cos(2\omega_{os}t)]e^{- \Gamma t},\hspace{5mm}
\end{eqnarray}
we see that the ``diagonal'' term that decays as $e^{-\Gamma t}$ is
second-order in the small amplitude fluctuations. The term linear in
$\Phi_{0}({\bf r})$ and linear in the fluctuations must be considered the
primary observable oscillating function. This term oscillates with a frequency
$\omega_{os}$ and decays as $e^{-\gamma t}$. It can be considered to arise
from a coherent superposition of the condensate and the condensate fluctuation
mode.  For our derivation to be relevant to this term, one must assume
that indeed $\gamma$ of Eq. (\ref{super}) is equal to $\gamma$ of
Eq. (\ref{transprob3}), {\it i.e.} that the ``off-diagonal'' element
decays at one-half the rate of the ``diagonal'' element.

One can question whether the decay always follows a simple exponential
law as we have assumed. For a more precise estimate of the time evolution of
the resonance oscillation, numerical integration of coupled equations,
including important off-diagonal density matrix elements such as the
coherent term in Eq. (\ref{super}), would be needed.

We note also that we have neglected the thermal atoms
in the $os$ mode, which will be significant at certain temperatures.
An accurate calculation of the contribution of this mode to the damping
process raises questions we have not resolved.  However, in view of the
additional constraint imposed by the Lorentz profile, we estimate that
Landau damping by the $os$ mode itself is not significant.

\section{Numerical Results for Excitation Widths}

To obtain the energies $\epsilon=\hbar\omega$ and the quasi-particle
wave-functions $u_{k}$ and the $v_{k}$, we solve self-consistently the
Gross-Pitaevskii equation for the condensate and the Bogoliubov equations,
as given in Eqs. (\ref{GPE}) and (\ref{Bog1}).  As stated above, here
we accept the Popov approximation and neglect the anomalous density
of the thermal particles, but always $n_{T}({\bf r})$ is included.
These equations are solved numerically using the discrete
variable representation (DVR) \cite{BayeHeenan,SF}, a method derived
from Gaussian quadrature methods. The sums over discrete quasi-particle
states are supplemented by local density approximation (LDA) expressions
\cite{Reidl}, which treat the high-lying states by a semiclassical
integral over momentum. More details of this part of the calculations
can be found in \cite{BFBS}.

To compute the widths we put in a trial value of $\gamma$ on the
right side of the implicit Eq. (\ref{maineq}) and evaluate the new
value of $\gamma$.  The calculation is iterated until the input
$\gamma$ is equal to the calculated $\gamma$ within a pre-defined
tolerance.  We made careful tests to ensure that the resultant
widths are independent of various computational parameters.
In particular, in evaluating the sum over states we
include all states such that $\omega^{\pm}_{ij}$ fell within a specified
range. For each mode studied, the range was increased until the width was
insensitive (within the tolerance) to the increments. Also we verified that
the width was insensitive to initial trial values.
Most of our calculations were done for
$\ell=0$ excitations. The set of perturbing levels ranged from $\ell=0$
up to $\ell_{max}$, ranging from 20 to 80, which was found sufficient to obtain
a converged value of $\gamma$.

\begin{figure}
\centering{\psfig{figure=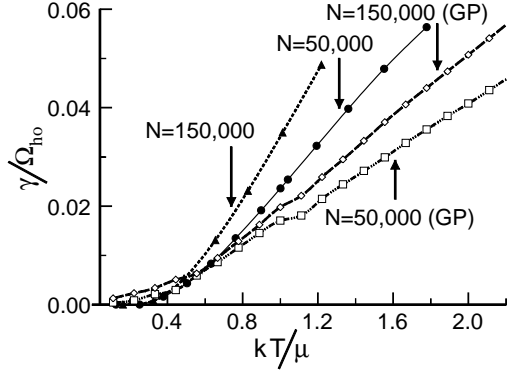,width=\columnwidth,angle=0}}
\caption{The decay width, $\gamma$, for the lowest energy mode
($n=1$), computed in this work as function of temperature is
compared to the widths computed by Guilleumas and Pitaevskii (GP)
in Ref. [30]. The width is scaled by the monopole resonance
frequency $\Omega_{{\rm ho}}\simeq\sqrt{5}\omega_{{\rm ho}}$ and
the temperature is scaled by the chemical potential, $\mu$, as in
this reference.} \label{compit}
\end{figure}

The above computational procedure is in principle more precise
than the methods used in Ref. \cite{GP}, which employ condensate and
excited state wavefunctions computed at $T=0$ with
populations appropriate to the stated $T$.  Although therefore we
do not expect exact agreement with these results, we made
extensive comparisons to provide qualitative tests of our approach.
For this purpose we used the same atom number values and
scaled scattering length, $a/r_{0} = 7.36 \times 10^{-3}$ as
in \cite{GP}. For example, the temperature variation of the widths
for $N$ = 50,000 and 150,000 from \cite{GP} and from our results
is shown in Fig. \ref{compit}.

\begin{figure}
\centering{\psfig{figure=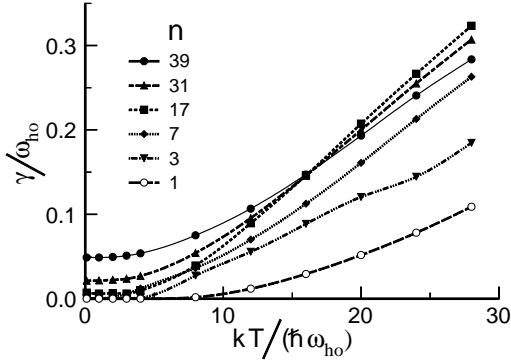,width=\columnwidth,angle=0}}
\caption{The decay width $\gamma$ as function of scaled
temperature, is shown for several condensate modes, for N=150,000
and $a/r_{0}=0.00736$.}\label{W150t}
\end{figure}

From our numerical calculations for a spherically symmetric
harmonic trap we observe the following trends:

\vspace{3mm}
(i) As in the data presented in \cite{GP}, we find
that the total width typically increases with temperature somewhat
linearly, as shown in Fig. \ref{W150t}. For the higher resonances,
non-zero values at $T$=0 reflect the Beliaev damping rate.

\begin{figure}
\centering{\psfig{figure=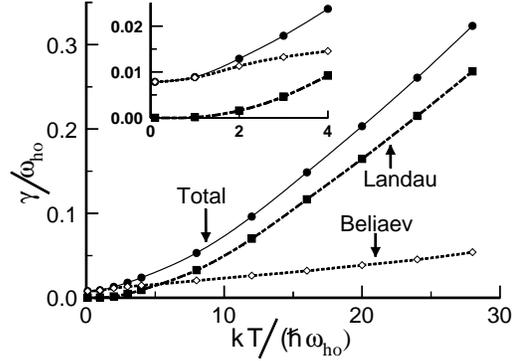,width=\columnwidth,angle=0}}
\caption{The total decay width and the Landau and Beliaev
contributions as function of scaled temperature
$kT/\hbar\omega_{{\rm ho}}$ are shown for the quasi-particle mode
with $n=15$, $N$=150,000 and $a/r_{0}=0.00736$.}\label{lbt_n15}
\end{figure}

\vspace{3mm}
(ii) Figure \ref{lbt_n15} explains that the temperature
increase of the width is due especially to the Landau
contribution, while the Beliaev part increases much more slowly.
This is plausible since the Landau process involves the
interaction of a thermally populated quasi-particle mode with the
oscillating mode in question, while the Beliaev process is simply
a decay into two modes of lower energy. At higher temperature, of
course there are more thermally populated modes available.

\begin{figure}
\centering{\psfig{figure=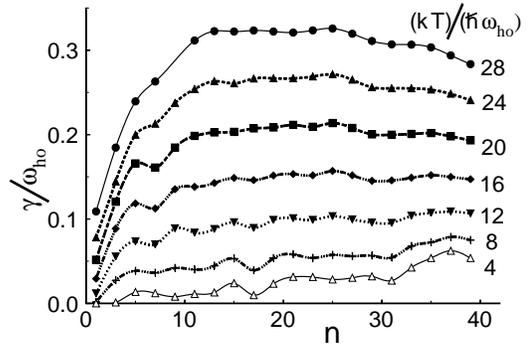,width=\columnwidth,angle=0}}
\caption{The decay width $\gamma$ as function of the number of
radial nodes, $n$, shown for a wide range of scaled temperatures,
for $N$=150,000 and $a/r_{0}=0.00736$.} \label{W150n}
\end{figure}

\vspace{3mm} (iii) The total width rises sharply
for low $n$, the radial quantum number, and then reaches a nearly flat
plateau at a value of $n$ roughly proportional to $T$, as shown in
Fig. \ref{W150n}. We have correlated
this effect with the factor $f_{j}^{0} - f_{i}^{0}$ in Eq. (\ref{transprob3}).
For the $(j,i)$ values that contribute most to $\gamma$, at higher
temperatures, $f_{j}^{0}$ is more nearly equal to $f_{i}^{0}$. The
temperature scale at which this occurs is set approximately by $(E_{i} -
E_{j})/k$. At higher $n$ values, $f_{i}^{0}$ is negligible compared with
$f_{j}^{0}$, so the width as a function of $n$ flattens out.
Note in this figure that
for low values of the temperature, the width is small compared
with $\hbar \omega_{ho}$ even for high values of $n$.

\begin{figure}
\centering{\psfig{figure=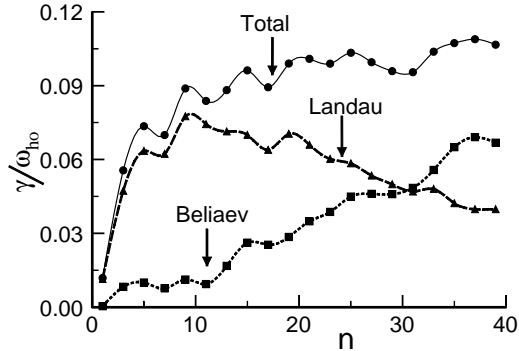,width=\columnwidth,angle=0}}
\caption{The total decay width and the Landau and Beliaev
contributions as function of the number of radial nodes $n$ of the
condensate fluctuation are shown for scaled temperature
$kT/\hbar\omega_{{\bf ho}}=12$, $N$=150,000 and
$a/r_{0}=0.00736$.}\label{lbn_t12}
\end{figure}

\vspace{3mm}
(iv) Figure \ref{lbn_t12}, again
breaks this trend down into Landau and Beliaev components for one
particular case. The apparent plateau occurs because of a near
balance between a decrease of the Landau rate and an increase in
the Beliaev rate. Since for higher energy, there are more final
states available, one does expect an increase in the Beliaev rate,
but it is surprising that the increase is as gradual as it is.
Also, the $n$ dependence of the Landau rate is rather unexpected.
We attribute the decrease at large $n$ to the decrease of the
overlap integrals in Eq. (\ref{pertpot2}), since the excited state
wavefunction will be increasingly extended in space.

\begin{figure}
\centering{\psfig{figure=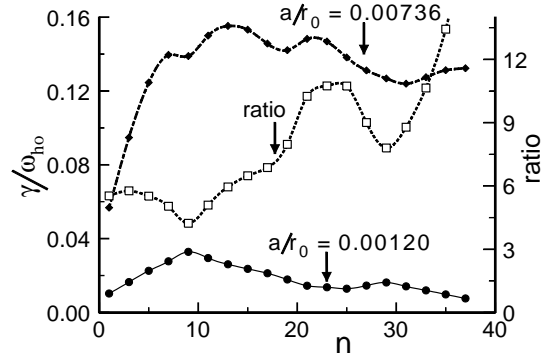,width=\columnwidth,angle=0}}
\caption{The decay width $\gamma$ as function of number of radial
nodes, $n$, of the condensate fluctuation, shown for two different
values of the scaled scattering length $a/r_{0}$. The data are for
$N$ = 50,000, and $kT/\hbar \omega_{{\rm ho}}$ = 16. Also shown is
the ratio of the widths (right axis), which may be compared with
the value of 6.13 for the ratio of the scattering
lengths.}\label{comsc_16}
\end{figure}

\vspace{3mm}
(v) The total width increases with scattering length in approximately
linear fashion, as shown for one set of conditions in
Fig. \ref{comsc_16}. This is consistent with Eq. (\ref{gamapp}). A
larger scattering length implies stronger
interaction among the particles and hence more rapid damping.

\begin{figure}
\centering{\psfig{figure=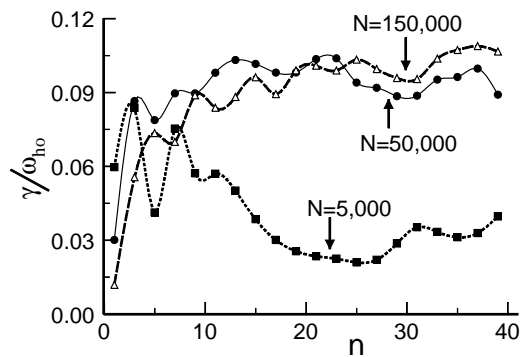,width=\columnwidth,angle=0}}
\caption{The decay width, $\gamma$, as function of number of
radial nodes, $n$, of the condensate mode, is shown for three
different values of the number of particles, N.  The data are for
scaled scattering length $a/r_{0}=0.00736$ and $kT/\hbar
\omega_{{\rm ho}} = 12$.}\label{com_n12}
\end{figure}

\vspace{3mm}
(vi) In Fig. \ref{com_n12}, we have plotted widths for
three different particle numbers.  With fewer particles,
the condensate is smaller and also the quasi-particle
wavefunctions are closer to bare harmonic oscillator functions.
For both reasons, the integrals in Eqs. (\ref{pertpot2}) yield smaller
values for the coupling parameters.

\vspace{3mm}
(vii) Especially for relatively low temperature and
smaller values of the scaled scattering length, the widths are
relatively small compared to the energy level separation, even at
high values of energy or $n$. Unfortunately, if one attempts to
excite high energy levels, typically many different $\ell$ values
would be excited and would overlap. However, if means can be found
to achieve a perfectly spherically symmetric harmonic trap and
then to excite a singlet $\ell$ manifold, our conclusion is that
the highly excited levels would be resolvable in this ideal case.

\vspace{3mm}
(viii) Although the above results were entirely for the case $\ell = 0$,
we have performed calculations for $\ell$ up to 6. For the cases
studied, have found only a slight variation of the width with $\ell$.

\section{Discussion and Conclusions}

A key question motivating this study has been the extent of validity of the
quasi-particle representation for trapped Bose atoms.
In light of recent kinetic theory results \cite{Stoof1,Stoof2}, it may be
that the quasi-particle representation is not essential for considering
resonance decay.  The appealing but complex picture of Landau and Beliaev
decay processes may not be necessary.  However, the Gross-Pitaevskii
equation and the Bogoliubov quasi-particle approach to Bose condensates
has been shown to be a powerful method for modeling thermal equilibrium
situations. Also in view of past successes \cite{Fedichev1,HDB},
it seems worthy of further study as well as application to more highly
excited resonances as we have done here. The dielectric formalism \cite{Reidl2}
employs quasi-particle basis states, but goes beyond the present
approximations.  We are currently analyzing these other
approaches to make comparisons.

It is useful to distinguish between discrete quasi-particle
modes, characterized by eigenvalues $\epsilon_{k}$ and eigenfunctions
$u_{k}({\bf r})$ and $v_{k}({\bf r})$ from Eq. (\ref{Bog1}), and semiclassical
quasi-particle functions, $u({\bf p,r}), v({\bf p,r})$ as used in \cite{GPS}.
The widths that we have computed for resonantly excited states may be
presumed to apply also to the thermal quasi-particle states, which must
thus be considered quasi-discrete quasi-particle (QDQP) states.  When the
widths of the QDQP states are comparable to their spacing, one can conclude
that discrete basis states are not relevant, and
a local semiclassical representation will suffice.  Spherical symmetry
presents a special case because the single-particle excitations at sufficiently
high energy within a given $\ell$ manifold have a spacing of $2 \hbar
\omega_{ho}$.  We do find that for $N = 150,000$ for example, for
$T < 12 \hbar \omega_{ho}/k$,
typical widths are $< 0.10 \hbar \omega_{ho}$, even for high-lying states
with $n$ up to at least 30. Going from spherical to cylindrical symmetry,
the spacing between levels within a given $m$ manifold decreases with energy,
and thus semiclassical quasi-particle functions are more likely to be accurate.
Consistent with this, we have observed that the semiclassical quasi-particle
functions become more reliable at lower energies in cylindrical than in
spherical symmetry for calculating thermal sums over quasi-particle states
\cite{BFBS,TB}.

Very recently, a harmonic trap close to spherical symmetry has been
produced \cite{EC1}, and we are informed \cite{EC2} that axial and transverse
frequencies can be made equal within about 1\%. Although we
have found above that resonances within a given $\ell$ manifold are under
certain circumstances resolvable, to our knowledge no one has yet devised an
excitation method that would be $\ell$ selective, as are electric dipole
excitations of atomic Rydberg levels.  One can hope that such techniques
for this purpose will be developed.
You and Walsworth \cite{Walsworth} have suggested that
time-varying magnetic fields be configured so as to excite
specific modes. With the help of focused laser beams, the MIT group has
observed resonances with $n_{z}$ as high as 8 in a cigar-shaped
trap, and surface modes \cite{Ketterle4} were excited by a sharply
focused perturbing laser beam, moved rapidly around the transverse
surface of the ``cigar'' trap.  The technique of stimulated Bragg
scattering \cite{Stenger,SKurn1,SKurn2} offers the capability to
select the excitation energy and wavenumber, but is not $\ell$
selective in its current form.  Possibly by special tailoring of
the laser beams, one could achieve a nearly spherically symmetric
excitation function.

Our review of the underlying theory of Bose condensate damping processes has
exposed some interesting contrasts with the Weisskopf-Wigner theory of
radiative damping.  The decay rates of Bose condensate excitations are
typically not a small fraction of the frequency, which makes the usual delta
function limit problematical.  Furthermore, the decay of interest is
very likely the ``off-diagonal'' density matrix element between the
condensate and the excited mode rather than the excitation intensity.
In the language of NMR, we have calculated
here $T_{1}$, but $T_{2}$ may be what is observed.  To model these various
circumstances in a more accurate manner, it may be necessary to explicitly
integrate coupled time-evolution equations for various diagonal and
off-diagonal elements.  Also, the assumption of simple exponential decay
may not always be valid.  Modulated decays have
been reported in calculations with an assumed 1D geometry \cite{Choi} and
also from Monte Carlo procedures \cite{Adams}. As is well known, a
system composed only of discrete frequencies will exhibit recurrences
rather than damping. This conclusion may not apply if the number of
interacting levels is effectively infinite, but at sufficiently low
temperature, there may {\it not} be effectively an infinite number of
accessible quasi-particle levels.  Again, more detailed modeling of
the time evolution may be needed.

In summary, we have presented numerical and perturbation theory
results for excitation frequencies of Bose particles in a
spherically symmetric harmonic trap, and compared with previously
derived expressions.  The energies approach the bare harmonic
oscillator values asymptotically  as $(4/3\pi)
\mu^{3/2}/\sqrt{2/n}$ independent of $\ell$ (for $n>>\ell$), where $n$
is the number of nodes in the radial equation. The asymptotic
regime is reached more slowly for higher $\ell$. Secondly, we have
presented a derivation of the excitation widths in agreement with
previous discussions, but based from the beginning on an assumption of
exponential decay, with an ensemble
of discrete quasi-particle states, hence yielding a Lorentzian
width expression rather than a delta function.  Thirdly, we have
presented numerical results for the widths over a range of
parameter values. They showing that the Landau process rises more rapidly with
temperature, while the Beliaev process rises more rapidly with
$n$, except for very low $n$. Also, compared with the frequency
separation of $2 \hbar \omega$ within an $\ell$ manifold, the
widths are often small, especially for low temperatures and
smaller values of the scaled scattering length.

This work was supported by ONR and NSF.  We gratefully acknowledge
valuable discussions with D. L. Feder, N. L. Balazs, B. I. Schneider,
H. J. Metcalf, A. Griffin, A. Kuklov and R. Walser.

\end{document}